# Effect of Strain on Ferroelectric Field Effect in Strongly Correlated Oxide $Sm_{0.5}Nd_{0.5}NiO_3$

L. Zhang,[1] X. G. Chen,[1] H. J. Gardner,[1] M. A. Koten,[2] J. E. Shield,[2,3] X. Hong[1,3,*]

[1] *Department of Physics and Astronomy, University of Nebraska-Lincoln, Lincoln, NE 68588 USA*

[2] *Department of Mechanical Engineering, University of Nebraska-Lincoln, Lincoln, NE 68588 USA*

[3] *Nebraska Center for Materials and Nanoscience, University of Nebraska-Lincoln, Lincoln, NE 68588 USA*

\* xia.hong@unl.edu

**Abstract**

We report the effect of epitaxial strain on the magnitude and retention of the ferroelectric field effect in high quality $PbZr_{0.3}Ti_{0.7}O_3$ (PZT)/3.8-4.3 nm $Sm_{0.5}Nd_{0.5}NiO_3$ (SNNO) heterostructures grown on (001) $LaAlO_3$ (LAO) and $SrTiO_3$ (STO) substrates. For SNNO on LAO, which exhibits a first-order metal-insulator transition (MIT), switching the polarization of PZT induces a 10 K shift in the transition temperature $T_{MI}$, with a maximum resistance change between the on and off states of $\Delta R/R_{on}$ ~75%. In sharp contrast, only up to 5% resistance change has been induced in SNNO on STO, where the MIT is second-order, with the modulation of $T_{MI}$ negligibly small. We also observe thermally activated retention of the off state resistance $R_{off}$ in both systems, with the activation energy of 22 meV (28 meV) for devices on LAO (STO). The time dynamics and thermal response of the field effect instability points to phonon-assisted interfacial trapping of charged mobile defects, which are attributed to strain induced oxygen vacancies. At room temperature, $R_{off}$ stabilizes at ~55% and ~19% of the initial switching levels for SNNO on LAO and STO, respectively, reflecting the significantly different oxygen vacancy densities in these two systems. Our results reveal the critical role of strain in engineering and modeling the complex oxide composite structures for nanoelectronic and spintronic applications.

Ferroelectric-strongly correlated oxide composite structures present a model system for examining quantum phase transitions driven by the competition between charge itineracy and various correlation effects due to strong electron-electron and electron-phonon interactions, as well as probing novel interface coupling mechanisms between different ferroic orders.[1, 2] In previous studies, ferroelectric field effect has been successfully utilized to modulate the electronic and magnetic properties of high-$T_c$ superconductors,[3-5] colossal magnetoresistive oxides,[6-14] and various Mott insulators.[15-17] This device concept also provides a viable route for building high density, low power, complex oxide-based logic and memory applications.[1-3] As many ferroelectric and correlated oxides belong to the perovskite family, it is desirable to combine them into epitaxial single crystalline heterostructures to achieve atomically sharp interfaces for low defect density and enhanced interface coupling.[1, 2] A central issue arising from the epitaxial growth is that strain imposed by the substrate can substantially alter the lattice parameters/structures of the parent crystals. As the electronic and magnetic properties of the correlated oxides are closely entangled with the lattice degree of freedom, it is critical to understand the role of strain to control and enhance the ferroelectric field effect in correlated oxides for nanoelectronic and spintronic applications.

In this Letter, we report a comprehensive study of the effect of epitaxial strain on the switching magnitude and retention characteristics of prototype ferroelectric field effect transistors based on ferroelectric Pb(Zr$_{0.3}$Ti$_{0.7}$)O$_3$ (PZT) and a charge-transfer type Mott insulator Sm$_{0.5}$Nd$_{0.5}$NiO$_3$ (SNNO). While switching the polarization field of PZT induces a 10 K shift in the metal-insulator transition (MIT) temperature $T_{MI}$ and a maximum resistance change between the high (off) and low (on) states of $\Delta R/R_{on}$ ~75% in samples grown on LaAlO$_3$ (LAO) substrates, it only yields an up to 5% resistance change in SNNO on SrTiO$_3$ (STO), with the shift of $T_{MI}$ negligibly small. In both systems, we observe thermally activated relaxation of the off state resistance, with the room temperature resistance stabilizing at ~55% and ~19% of the initial switching levels for devices on LAO and STO, respectively. We believe the ferroelectric field effect instability is due to partial polarization screening induced by a slow interfacial charge trapping process, and a likely source of the charged mobile defects is strain induced oxygen vacancy, as suggested by the time dynamics and thermal response of the retention behavior. Our results highlight the critical role of strain in determining the performance of these complex oxide composite devices.

We deposited epitaxial PZT/SNNO heterostructures on (001) LAO and STO substrates using off-axis radio frequency magnetron sputtering. Bulk SNNO has a pseudocubic lattice constant of 3.804 Å,[18] subjecting to a 2.7% (tensile) strain on STO and a -0.37% (compressive) strain on LAO. The growth conditions for SNNO can be found in Ref. [19]. A 90-150 nm PZT layer was then deposited *in situ* on SNNO at 520˚C, with 150 mtorr process gas composed of Ar and $O_2$ (ratio of 2:1). After growth, the samples were cooled down in 1 atmosphere of $O_2$. These samples show smooth surface morphology with a typical root mean square surface roughness of ~5 Å (Figs. 1(a) insets). X-ray $\theta$-$2\theta$ scans reveal predominately *c*-axis growth for PZT with the out-of-plane lattice constant of ~4.12 Å (Fig. 1(a)). The high-resolution transmission electron microscopy (HRTEM) measurements reveal atomically sharp interfaces in these heterostructures (Fig. 1(b)). The as-grown PZT film is uniformly polarized in the down orientation ($P_{down}$), as characterized by piezo-response force microscopy (PFM) (Fig. 1(c)). The heterostructures were then fabricated into four-point devices with a top gate using optical lithography followed by Au deposition as contact electrodes (Fig. 1(d)). The transport measurements were performed using

Quantum Design Physical Property Measurement System combined with Keithley 2400 Source Meters. We used excitation current below 1 µA to avoid Joule heating.

Figure 1(e) shows the sheet resistance $R_\square$ as a function of gate voltage $V_g$ for a 4 nm SNNO device on LAO at 300 K. As SNNO possesses $p$-type charge carriers,[19] the polarization up $P_{up}$ ($P_{down}$) state corresponds to the accumulation (depletion) of holes in the channel, yielding a low (high) channel resistance value. The resistance switching occurs at about ±2 V, agreeing with the coercive field of PZT at room temperature. The switching voltage increases with decreasing temperature, reaching about ±6V at 10 K, which is attributed to the suppressed switching kinetics associated with domain nucleation and growth at low temperature.[20, 21] Figure 1(f) shows reliable resistance switching between the on and off states for a 4 nm SNNO device on LAO upon the application of a train of alternating ±5 V pulses with 1s pulse width across PZT.

Bulk perovskite nickelate $R$NiO$_3$ ($R$ = rare earth except La) exhibits a temperature-driven MIT as well as a paramagnetic to antiferromagnetic (AFM) transition, with both depending sensitively on lattice distortion.[22, 23] Significant modifications to their phase transition characteristics and $T_{MI}$ can be achieved by applying hydrostatic pressure,[22, 23] imposing epitaxial strain,[19, 24-28] or varying internal chemical pressure via $R$-site cation substitution.[29, 30] In particular, SNNO is close to the structural phase boundary where the $T_{MI}$ is about to decouple from the AFM Neel temperature and the MIT evolves from first-order to second-order, and thus exhibits high susceptibility to the structural modifications. Figure 2(a) shows $R_\square(T)$ upon warming for a 4 nm SNNO on LAO for both polarization states. This film exhibits a pronounced $R(T)$ hysteresis in the heating-cooling cycle (Fig. 2(a) inset), signaling a first-order transition,[23] qualitatively similar to those observed on single layer SNNO of similar thickness on LAO,[19] while the MIT in bulk SNNO is second-order.[22] The $P_{up}$ state accumulates holes in SNNO, and the corresponding $T_{MI}$, where d$R$/d$T$ changes sign upon warming, is 220 K. The resistance rises sharply with decreasing temperature between 170 K and 120 K, which has been attributed to percolative transport due to the co-existing semiconducting and metallic phases. At low temperature, the transport can be well described by the 2D variable range hopping (VRH) model (Fig. 2(c)):[31]

$$\sigma\sqrt{T} = \sigma_0 \left[-\left(\frac{T_0}{T}\right)^{1/3}\right], \quad \text{where} \quad T_0 = \frac{13.8}{k_B N_{2D}(E_F)a^2} . \qquad (1)$$

Here $a$ is the localization length of the charge carriers, and $N_{2D}(E_F)$ is the 2D density of impurity states (DoS) at the Fermi level, which is estimated based on the 3D DoS $N_{3D}(E_F) = 6.2\times10^{18}$ eV$^{-1}$ cm$^{-3}$ extracted from thicker SNNO films[19] and the film thickness. Fitting the low temperature data to Eq. (1) yields a localization length of 69 nm. At 17 K, the hopping energy $W = \frac{1}{\pi R^2 N_{2D}(E_F)} = 1.3$ meV becomes comparable to the thermal energy $k_B T$, and $\sigma(T)$ starts to deviate from the 2D VRH model (Fig. 2(c)).

We then applied a voltage pulse of +10 V across PZT at 10 K to switch it to the $P_{down}$ state, which depletes holes from SNNO. As expected, the film becomes more resistive, and $T_{MI}$ increases to 230 K (Fig. 2(b)). At low temperature, fitting $\sigma(T)$ to Eq. (1) yields a localization length of 70 nm (Fig. 2(c)), identical to the value obtained in the $P_{up}$ state within the error range. This clearly indicates that the polarization switching does not modify the degree of charge localization despite an up to 50% resistance modulation in this temperature range. It has been shown that charge localization in nickelates is strongly influenced by lattice distortion,[19] with the Ni-O-Ni bond angle and the Ni-O bond length collectively determining the itineracy of the charge carriers. This result thus suggests strongly that the resistance change originates from the electrostatic tuning of the Fermi level rather than structural modification of the lattice. From Hall effect measurements at 225 K, close to the metallic phase, we extracted a carrier density of $7.29\times10^{22}$/cm$^3$ for the $P_{up}$ state and $7.02\times10^{22}$/cm$^3$ for the $P_{down}$ state, consistent with the values obtained on thicker films.[19] The density modulation corresponds to a polarization field of 88 µC/cm$^2$, agreeing with the expected value for PZT.

The magnitude of the field effect is closely related to the MIT characteristic of SNNO. Figure 2(b) shows the temperature dependence of resistance switching ratio $\Delta R/R_{on}$ for this device, where $\Delta R = R_{off} - R_{on}$. In the metallic phase at 250 K, $\Delta R/R_{on}$ is only 4%, consistent with the level of carrier density modulation. As the system enters the percolative transport regime, the $R(n)$ relationship becomes highly nonlinear, which results in a non-monotonic temperature dependence of $\Delta R/R_{on}$ with a maximum resistance modulation of ~75% at 140 K. Below 100 K, the system is well residing in the insulating phase, and $\Delta R/R_{on}$ shows moderate increase with decreasing temperature, reaching ~50% and 25 K.

In sharp contrast, SNNO on STO exhibits a broad, second-order MIT with no apparent thermal hysteresis (Fig. 2(d)). Charge carriers in this device is highly localized, leading to a higher $T_{MI}$ of ~335 K, similar to what is observed on single layer SNNO of similar thickness on STO.[19] Assuming the same carrier density,[19] we obtained a high temperature mobility of 0.039 cm$^2$ V$^{-1}$ s$^{-1}$, while the mobility for the device on LAO is 0.11 cm$^2$ V$^{-1}$ s$^{-1}$. The significantly lower mobility for SNNO on STO has been attributed to the large tensile strain, which stretches the Ni-O bond length, resulting in softened oxygen vibration modes that enhances charge localization.[19,22] This effect, combined with a suppressed correlation gap due to the cubic symmetry imposed by the STO substrate, leads to a weaker temperature dependence of resistance in the intermediate temperature regime.[19] A direct consequence is that switching the polarization to the $P_{down}$ state does not yield pronounced modulation in the resistance, with negligibly small shift in $T_{MI}$. Between 350 K and 120 K, $\Delta R/R_{on}$ is about 1-2% showing no apparent temperature dependence (Fig. 2(e)). Below 120 K, $\Delta R/R_{on}$ increases with decreasing temperature, and reaches 5% at 75

K. The drastic difference in the magnitude of field effect modulation in SNNO on different substrates is a direct manifestation of the nonlinear resistance-density relation for the strongly correlated systems. The maximum resistance change is observed in the highly inhomogeneous phase for SNNO on LAO, which is absent for the system on STO as the MIT is second-order.

The magnitude of the field effect modulation is further attenuated by a thermally activated relaxation behavior for the off state resistance. Figure 2(f) shows the retention of the on and off states for SNNO on LAO and STO at 200 K. While $R_{on}$ is stable within 5% for both systems, $R_{off}$ gradually decays to about 70% and 35% of the initial switching levels for devices on LAO and STO, respectively. The residue resistance level decreases with increasing temperature, and the high temperature decay is more pronounced in devices on STO (Fig. 3(a)-(b)).

A possible origin for the dynamic response of $R_{off}$ is the existence of a slow interfacial charging mechanism triggered by the polarization reversal. As the electric field effect is resulting from the screening of the polarization field via redistributing the carrier density in the conducting channel, any additional interfacial screening mechanism can undermine the field effect sensed by SNNO. To identify the source of the interfacial charges, we quantitatively modeled the retention behavior of $R_{off}$. As shown in Fig. 3(c)-(d), the time dependence of normalized resistance can be well described by a stretched exponential model:[32]

$$R_{norm}(t) = (1 - R_f)exp\left[-\left(\frac{t}{\tau}\right)^\beta\right] + R_f. \qquad (2)$$

Here $R_f$ is the residual percentage as $t \rightarrow \infty$, $\tau$ is the characteristic relaxation time constant, and $\beta$ is the stretching exponent that should be between 0 and 1. Such type of time response has been widely observed in disordered materials such as amorphous semiconductors[32, 33] and polymers,[34] and has been attributed to dispersive transport of mobile defects. For the device on STO, we also observed a rapid exponential decay of $R_{off}$ at $t < 100$ s (Fig. 3(d)). This is likely due to the strong depolarization field in this system, as the charge carriers in SNNO on STO are more localized and cannot provide sufficient polarization screening.[21]

By fitting the retention data to Eq. (2), we found the relaxation time $\tau$ to be ~4.4×10³ s and 1.4×10² s at 300 K for the devices on LAO and STO, respectively, and decreases exponentially with increasing temperature (Fig. 4(a)). Using a thermal activation model $\tau = \tau_0 \exp(E_a/k_B T)$, we extracted the activation energy $E_a$ of the mobile defects to be 22±1 meV for SNNO on LAO and 28±3 meV for SNNO on STO.

It has been shown that the temperature dependence of the stretching exponent $\beta$ can reveal critical information on the nature of the mobile defects.[32-34] As shown in Fig. 4(b), for both systems, $\beta(T)$ can be well fitted using a hyperbolic tangent function, as proposed by Apitz and Johansen for disordered polymeric systems that exhibit a glass transition:[35]

$$\beta(T) = \frac{\beta_\infty}{2}\left[1 - tanh\left(\frac{E'_a - \alpha T}{k_B T}\right)\right]. \quad (3)$$

Such a temperature dependence can be understood as an ensemble average of the charge trapping and de-trapping (releasing) processes, where $\beta_\infty$ is the high temperature upper limit for $\beta$, and $E'_a$ is the activation energy of mobile defects at $T = 0$. We extracted the trapping energies $E'_a$ for the charged defects to be 16±3 meV and 29±5 meV for devices on LAO and STO, respectively, comparable to the activation energies obtained from $\tau(T)$. At finite temperature, the trapping energy is reduced by a term that is proportional to temperature, suggesting a possible contribution of phonon-assisted hopping, and the $\alpha$ coefficients are 0.20 meV/K and 0.24 meV/K for devices on LAO and STO, respectively. At high temperature, $\beta$ for both systems approaches $\beta_\infty$ ~0.4 and shows very weak temperature dependence. It is worth noting that the deflection point $T_0$ in $\beta(T)$, which is 80 K (118 K) for the device on LAO (STO), is exactly the temperature at which the reduced trapping energy becomes comparable with the thermal energy, i.e., $E'_a - \alpha T_0 \sim k_B T_0$. The trapped defect states become delocalized above $T_0$, leading to hopping transport that depends weakly on temperature.[32] This is consistent with what is observed in disordered polymeric systems, where $T_0$ coincides with the glass transition temperature.[34]

The fact that the field effect instability occurs mainly for the off state points strongly to negatively charged mobile defects. Upon the switching of the polarization to the $P_{down}$ orientation, the negative charges slowly migrate to the interface, providing effectively partial screening of the polarization field. As a result, the charge carriers in the nickelates would only respond to a reduced polarization field. A likely source of mobile defects in epitaxial nickelates is oxygen vacancies resulting from strain relaxation. It has been shown that oxygen vacancies have profound impact on the transition characteristic in nickelates.[28, 36] For SNNO thin films on STO, the density of defect states at the Fermi level is three orders of magnitude larger than that of SNNO on LAO due to the substantial tensile strain,[19] which can lower the oxygen vacancy formation energy.[37] This high defect density can account for the more pronounced high temperature relaxation in devices on STO. As shown in Fig. 4(b) inset, while $R_{off}$ for both systems stabilizes at around 75% at 100 K, $R_f$ is only 19% at 300 K for the device on STO, in contrast to 55% for the device on LAO.

In summary, using PZT/SNNO heterostructures as a model system, we have systematically evaluated the role of epitaxial strain in the ferroelectric field effect modulation of strongly correlated oxides. We find that the magnitude of the resistance change is closely related to the MIT characteristics in SNNO, with the maximum modulation observed in the percolative transport regime for devices on LAO. Both systems exhibit thermally activated retention behavior for the off state resistance, which has been attributed to partial polarization screening induced by interfacial trapping of mobile defects. We believe the charged defects originate from strain induced oxygen vacancies. Our results provide important insights into the role of epitaxial strain for the functional design and modeling of these composite oxide electronic devices.

We would like to thank Vijay Raj Singh, Zhiyong Xiao, and Xiaozhe Zhang for technical assistance. The sample growth and structural characterizations have been supported by the NSF Nebraska Materials Research Science and Engineering Center (MRSEC) (Grant No. DMR-1420645). The device fabrication and transport studies have been supported by the Center for NanoFerroic Devices (CNFD) and the Nanoelectronics Research Initiative (NRI). The data analysis and manuscript preparation have been supported by NSF Grant No. DMR-1409622 and NSF CAREER Grant No. DMR-1148783. This research was performed in part in Central Facilities of the Nebraska Center for Materials and Nanoscience, which is supported by the Nebraska Research Initiative.

**Captions:**

FIG. 1. (a) X-ray $\theta$-$2\theta$ scans for PZT/SNNO on LAO and STO show predominately *c*-axis growth with a small fraction of *a*-axis of PZT. Insets: AFM topography of the samples. (b) Cross-sectional HRTEM image of a PZT/SNNO (10 unit cell)/LAO heterostructure along [110]. The dotted lines mark the interfaces. (c) PFM phase image of a PZT/ SNNO/STO heterostructure. The center area was polarized to the $P_{up}$ ($P_{down}$) state by applying -3 V (+3 V) to the AFM tip. (d) Device schematic. (e) $R_\square(V_g)$ hysteresis taken on a 4 nm SNNO device on LAO. (f) Resistance switching cycles taken on a device with 4 nm SNNO on LAO.

FIG. 2. (a) $R_\square(T)$ taken on a 4 nm SNNO on LAO for the $P_{up}$ (blue) and $P_{down}$ (red) states of PZT. The arrows mark the corresponding $T_{MI}$s. Inset: the heating/cooling $R_\square(T)$ hysteresis for the $P_{up}$ state. (b) $\Delta R/R_{on}$ $(T)$ and (c) 2D VRH fits to the low temperature transport for the device in (a). (d) $R_\square(T)$ for the $P_{up}$ (blue) and $P_{down}$ (red) states and (e) $\Delta R/R_{on}$ $(T)$ for a 4.3 nm SNNO on STO. (f) Retention studies of the on ($P_{up}$) and off ($P_{down}$) states for the devices on LAO (dashed lines) and STO (solid lines) at 200 K.

FIG. 3. The retention behaviors of normalized $R_{off}$ at various temperatures for (a) a 4 nm SNNO on LAO and (b) a 4.3 nm SNNO on STO. (c)-(d) The log-log plots of the retention data (symbols) in (a) and (b), respectively, with fits to Eq. (2) (solid lines). The dashed lines in (d) are fits to the exponential decay model $\exp(-t/\tau')$ with $\tau'$ of $2.0\times10^2$ s at 300 K and $2.5\times10^3$ s at 100 K.

FIG. 4. Extracted parameters for $R_{off}$ from Eq. (2) for the devices on LAO (red square) and STO (blue circle): (a) $\tau(T)$ with fits to the thermal activation model (dotted lines), (b) $\beta(T)$ with fits (dotted lines) to Eq. (3), and (inset) $R_f(T)$.

# Figure 1

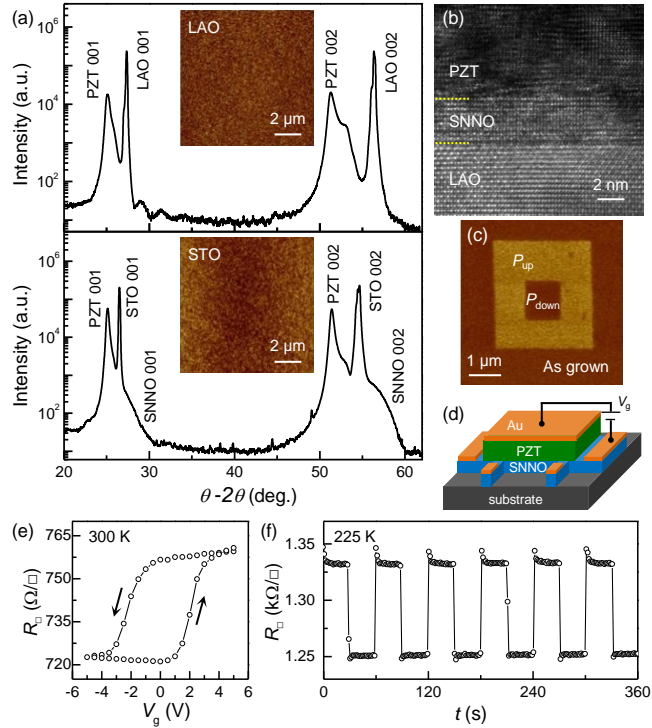

FIG. 1. (a) X-ray $\theta$-$2\theta$ scans for PZT/SNNO on LAO (top) and STO (bottom) show predominately *c*-axis growth with a small fraction of *a*-axis of PZT. Insets: AFM topography of the samples. (b) Cross-sectional HRTEM image of a PZT/SNNO (10 unit cell)/LAO heterostructure along [110]. The dotted lines mark the interfaces. (c) PFM phase image of a PZT/SNNO/STO heterostructure. The center area was polarized to the $P_{up}$ ($P_{down}$) state by applying -3 V (+3 V) to the AFM tip. (d) Device schematic. (e) $R_\Box(V_g)$ hysteresis taken on a device with 4 nm SNNO on LAO. (f) Resistance switching cycles taken on a device with 4 nm SNNO on LAO.

Figure 2

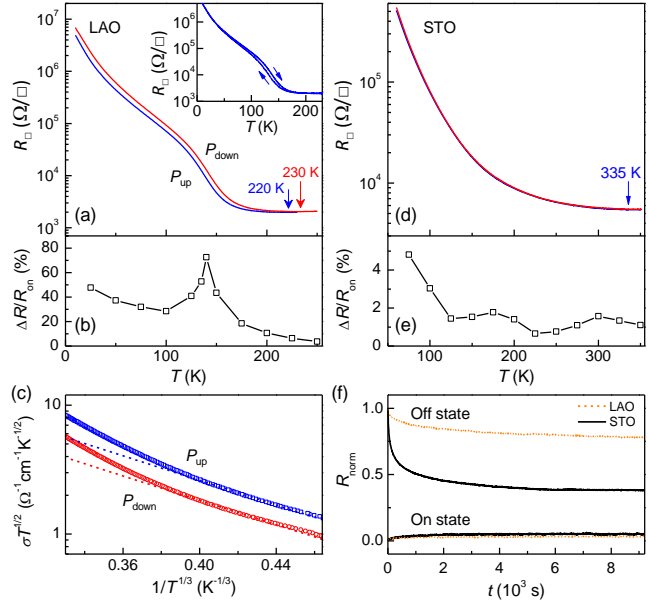

FIG. 2. (a) $R_\square(T)$ taken on a 4 nm SNNO on LAO for the $P_{up}$ (blue) and $P_{down}$ (red) states of PZT. The arrows mark the corresponding $T_{MI}$s. Inset: the heating/cooling $R_\square(T)$ hysteresis for the $P_{up}$ state. (b) $\Delta R/R_{on}(T)$ and (c) 2D VRH fits to the low temperature transport for the device in (a). (d) $R_\square(T)$ taken on a 4.3 nm SNNO on STO for the $P_{up}$ (blue) and $P_{down}$ (red) states of PZT. The shift in $R(T)$ is negligibly small. (e) $\Delta R/R_{on}(T)$ for the device in (d). (f) Retention studies of the on ($P_{up}$) and off ($P_{down}$) states for the devices on LAO (dashed lines) and STO (solid lines) at 200 K.

# Figure 3

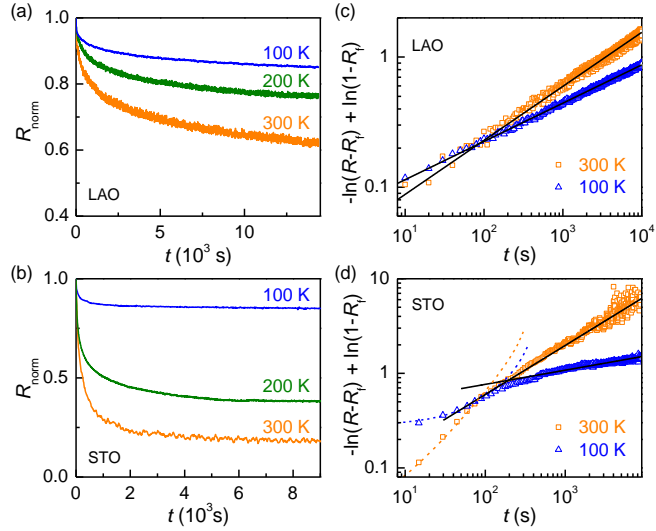

FIG. 3. The retention behaviors of normalized $R_{off}$ at various temperatures for (a) a 4 nm SNNO on LAO and (b) a 4.3 nm SNNO on STO. (c)-(d) The log-log plots of the retention data (symbols) in (a) and (b), respectively, with fits to Eq. (2) (solid lines). The dashed lines in (d) are fits to the exponential decay model $\exp(-t/\tau')$ with $\tau'$ of $2.0\times10^2$ s at 300 K and $2.5\times10^3$ s at 100 K.

Figure 4

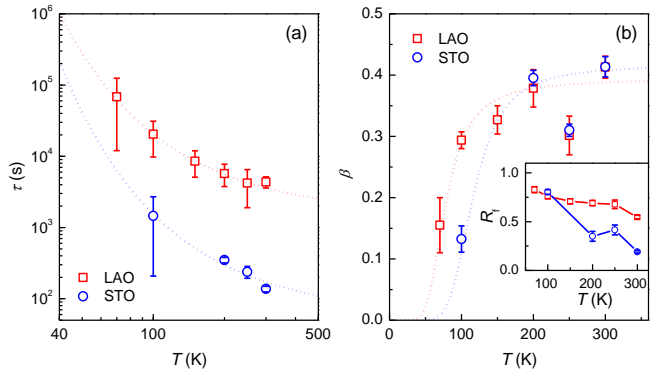

FIG. 4. Extracted parameters for $R_{off}$ from Eq. (2) for the devices on LAO (red square) and STO (blue circle): (a) $\tau(T)$ with fits to the thermal activation model (dotted lines). (b) $\beta(T)$ with fits (dotted lines) to Eq. (3), and (inset) $R_f(T)$.